\documentclass[12pt]{article} 
\usepackage{graphicx}
\usepackage{epsfig}
\usepackage{amsbsy}

\begin{document}
\baselineskip 10mm

\centerline{\large \bf Formation of a {}``Cluster Molecule'' (C$_{20}$)$_{2}$}
\centerline{\large \bf and Its Thermal Stability}

\vskip 6mm

\centerline{A. I. Podlivaev and L. A. Openov$^{*}$}

\vskip 4mm

\centerline{\it Moscow Engineering Physics Institute (State
University), 115409 Moscow, Russia}

\vskip 2mm

$^{*}$ E-mail: LAOpenov@mephi.ru

\vskip 8mm

\centerline{\bf ABSTRACT}

The possible formation of a {}``cluster molecule'' (C$_{20}$)$_{2}$ from two single C$_{20}$
fullerenes is studied by the tight-binding method. Several (C$_{20}$)$_{2}$ isomers in which C$_{20}$
fullerenes are bound by strong covalent forces and retain their identity are found; actually, these C$_{20}$
fullerenes play the role of {}``atoms'' in the {}``cluster molecule''. The so-called {\it open}-[2+2] isomer has a minimum energy. Its formation
path and thermal stability at $T=2000\div 4000$ K are analyzed in detail. This isomer loses its molecular structure
due to either the decay of one of C$_{20}$ fullerenes or the coalescence of two C$_{20}$
fullerenes into a C$_{40}$ cluster. The energy barriers for the metastable {\it open}-[2+2] configuration are
calculated to be $U=2\div 5$ eV.

\newpage

\centerline{\bf 1. INTRODUCTION}

The discovery of fullerene C$_{60}$ [1] stimulated extensive theoretical and experimental studies of carbon
clusters and other carbon nanostructures. In 2000, the fullerene C$_{20}$ that is the smallest among all
possible fullerenes was detected (Fig. 1): on its surface, C-C bonds form only regular pentagons and there are
no hexagons (unlike in C$_{n}$ fullerenes with  $n\,>\,20$) [2]. Later, the authors of [3] obtained experimental
evidence for the charged cluster dimers and (C$_{20}$)$_k^+$ with $k= 3\div 13$. In the future, it would be
interesting to synthesize a macroscopic C$_{20}$-fullerene-based cluster matter (by analogy with a fullerite
made of C$_{60}$ clusters [4]). According to theoretical studies [5-7], such a matter can be a superconductor
with an extremely high critical temperature.

An analysis of the paths of formation of the C$_{20}$ fullerite should begin with a detailed discussion of C$_{20}$
cluster dimerization. The structural and energetic characteristics of  (C$_{20}$)$_{2}$ dimers were first studied
theoretically in [5] (by the density functional method in the local-density approximation with gradient corrections)
and in [8] (by the Hartree-Fock and density functional methods with the B3LYP exchange-correlation functional).
The authors of [8] found several metastable (C$_{20}$)$_{2}$ isomers with different number, strength, and
length of intercluster bonds. Three of them are shown in Fig. 2. The {\it open}-[2+2] isomer has a minimum
energy (Fig. 2c); hereafter numerals in square brackets mean the number of atoms of each fullerene involved in
the intercluster bonding. As was shown in [8], when the {\it open}-[2+2] isomer forms, the energy of
the (C$_{20}$)$_{2}$ system remains below the total energy of two single C$_{20}$ fullerenes approaching each
other; that is, the open {\it open}-[2+2] configuration is favorable from both the kinetic and thermodynamic viewpoint.

In this work, the path of formation of a {}``cluster molecule'' (C$_{20}$)$_{2}$ from two
C$_{20}$ fullerenes is studied using the tight-binding potential [9]. We analyze the potential-energy surfaces at
various sections of this path and find the energy barriers that hinder the loss of molecular structure of the
(C$_{20}$)$_{2}$ dimer. Using molecular dynamics simulation, we investigate the evolution of the
(C$_{20}$)$_{2}$ dimer in real time at temperatures high enough to overcome the barriers. The results obtained
indicate that the {\it open}-[2+2] isomer is rather stable.

\vskip 5mm

\centerline{\bf 2. COMPUTATIONAL PROCEDURE}

We calculate the potential energy of the (C$_{20}$)$_{2}$ system
by the tight-binding method with the transferable interatomic
potential [9] in the Born-Oppenheimer approximation at fixed
coordinates $\left\{ \mathbf{\mathrm{\mathbf{R}}}_{i}\right\}$ of
all atoms ($i=1-40$ is the atom number). This method is a
reasonable compromise between oversimplified classical approaches
and {\it ab initio} calculations requiring much computational
time. Earlier [10-14], we used this method to simulate various
carbon clusters, including C$_{60}$ and C$_{20}$ fullerenes. The
energy $E_{pot}(\left\{ \mathbf{\mathrm{\mathbf{R}}}_{i}\right\})$
is equal to the sum of the classical atomic repulsion energy and
the so-called band energy, which can be found by diagonalizing the
Hamiltonian matrix in the site representation and by summation the
energies of one-electron levels occupied according to the Pauli
principle [9]. We took into account the valence electrons
occupying the 2$s$, 2$p_x$, 2$p_y$, and 2$p_z$ orbitals of each
carbon atom. The interatomic distances, binding energies,
HOMO-LUMO gaps, and other characteristics calculated for C$_{60}$
and C$_{20}$ clusters using this method agree with the
experimental data and the results of {\it ab initio} calculations
[13, 14].

The forces $\mathbf{F}_{i}$ acting on atoms were determined from the Hellmann-Feynman formula by calculating
the matrix elements of the gradient of the tight-binding Hamiltonian between occupied eigenstates.
The temperature $T_{el}$ of the electron subsystem was taken to be zero, which simplifies the calculations.
We chose this approximation due to the following reasons: first, the results of molecular dynamics for a single
C$_{20}$ fullerene at $T_{el}=0$ and 3000 K differ only slightly [13] and, second, in the
(C$_{20}$)$_{2}$ dimer, the HOMO-LUMO gap separating the upper unoccupied one-electron orbitals from the lower
occupied ones is larger than that in the C$_{20}$ fullerene (0.65 and 0.43 eV, respectively). As a result, excited electron
states are less significant in the temperature range under study.

To study the evolution of the (C$_{20}$)$_{2}$ dimer at high temperatures, we performed molecular-dynamics
simulation with the tight-binding potential used in [9] and a time step $t_0=2.72\cdot 10^{-16}$ s. The total energy
of the system (the sum of the potential and kinetic energies) remained unchanged during simulation, and the system
temperature $T$ was determined from the formula
\begin{equation} \frac{1}{2}k_{B}T(3n-6)=\langle
E_{\mathrm{kin}}\rangle  ,
\end{equation}
where $k_B$ is the Boltzmann constant, $n=40$ is the number of atoms in the system, and
$\langle E_{\mathrm{kin}}\rangle$ is the time-averaged kinetic energy. This formulation of the problem corresponds
to the situation where the system is not in thermal equilibrium with its environment. The microcanonical temperature $T$
is a measure of the energy of relative atomic motion [15]. At the initial time, all atoms were given random velocities
and displacements such that the momentum and the angular momentum of the system were zero. Then, we calculated
the forces acting on atoms and numerically solved the classical Newton equations of motion.

To analyze the potential-energy surfaces $E_{\mathrm{pot}}\left(\left\{ \mathbf{R}_{i}\right\} \right)$,
to determine the paths of the system between various states, and to find the heights $U$ of the energy barriers
present in these paths, we performed structure relaxation and found the saddle points of the energy
$E_{\mathrm{pot}}\left(\left\{ \mathbf{R}_{i}\right\} \right)$ as a function of the normal coordinates
corresponding to unstable atomic equilibrium [13].

\vskip 5mm

\centerline{\bf 3. FORMATION OF THE (C$_{20}$)$_{2}$ DIMER}

As in [8], we found several metastable configurations
(C$_{20}$)$_{2}$. Three of them, including the {\it open}-[2 + 2]
isomer (which has a minimum energy), are shown in Fig. 2. The
binding energies $E_b$ of these isomers were calculated from the
formula
\begin{equation}
E_{b}\,=\,E_{\mathrm{pot}}\left(\mathrm{C}_{1}\right)-E_{\mathrm{pot}}\left(\mathrm{C}_{n}\right)/n,
\end{equation}
where $n$ is the number of atoms in the system ($n=40$ for the
(C$_{20}$)$_{2}$ dimer), $E_{pot}$(C1) is the energy of a single
carbon atom, and $E_{pot}(C_n)$ is the energy of an $n$-atom
configuration. The binding energies were calculated to be 6.14,
6.16, and 6.20 eV/atom for the [1 + 1], [2 + 2], and {\it open}-[2
+ 2] isomers, respectively. All of these values of $E_b$ are
higher than the binding energy of one C$_{20}$ fullerene
($E_b=6.08$ eV/atom) calculated by the same tight-binding method
[13]. Therefore, the energy $E_{pot}$ of these isomers is lower
than the total energy of two single C$_{20}$ fullerenes; that is,
their formation is favorable from a thermodynamic viewpoint. The
coalescence energies $\Delta
E\,=\,2E_{\mathrm{pot}}\left[\mathrm{C}_{20}\right]-E_{\mathrm{pot}}\left[(\mathrm{C}_{20})_{2}\right]$
for the [1 + 1], [2 + 2], and {\it open}-[2 + 2] isomers are 2.5,
3.2, and 4.9 eV, respectively. The values of $\Delta E$ calculated
by the Hartree-Fock and density functional methods are 2.3, 5.9,
and 7.1 eV and 2.4, 4.7, and 6.3 eV, respectively [8]. It is seen
that, although the absolute values of $\Delta E$ differ quite
significantly, different theoretical approaches result in the same
sequence of (C$_{20}$)$_{2}$ isomers from an energetic viewpoint.
The validity of our potential is also supported by the fact that
the bond lengths in (C$_{20}$)$_{2}$ dimers agree well with the data
from [8] (see Fig. 2).

An analysis of the potential energy $E_{pot}$ as a function of the atomic coordinates
$\left\{ \mathbf{\mathrm{\mathbf{R}}}_{i}\right\}$ demonstrates that there is no barrier in the path of formation of
the [1 + 1] isomer from two C$_{20}$ fullerenes (Fig. 3a). As a result of the coalescence of two fullerenes, a
C$_{20}$)$_{2}$ dimer with one 1-2 bond forms (numerals indicate the atom numbers; see Fig. 2). In turn, the
[1 + 1] isomer can transform into a [2 + 2] isomer, which has a lower energy, by overcoming a barrier $U=0.33$ eV
(Fig. 3b). In this case, the C$_{20}$ fullerenes rotate about the 1-2 bond with respect to each other, which leads to the
formation of a second (3-4) bond between them. The barrier height for the [2 + 2] isomer to transform into the
{\it open}-[2 + 2] isomer (Fig. 2c), which has a minimum energy, is $U=0.57$ eV (Fig. 3c). This transition proceeds
via the sequential breaking of intracluster 1-3 and 2-4 bonds.

Our results are in overall agreement with the results of [8] except for the fact that the authors of [8] failed to
determine the barrier height in the [1 + 1] $\rightarrow$ [2 + 2] path because of the small curvature of the potential
surface in the vicinity of the [1 + 1] configuration. Nevertheless, we confirmed the main conclusion drawn in [8] that,
along the entire path of the C$_{20}$ + C$_{20}$ $\rightarrow$ [1 + 1] $\rightarrow$ [2 + 2] $\rightarrow$
{\it open}-[2 + 2] transition (including saddle points), the energy of the system remains below the total energy of two
C$_{20}$ fullerenes located far from each other (Fig. 3). Hence, the formation of the {\it open}-[2 + 2] isomer does
not require energy consumption and is favorable from both the thermodynamic and kinetic viewpoint.

\vskip 5mm

\centerline{\bf 4. STABILITY OF THE (C$_{20}$)$_{2}$ DIMER}

Although the {\it open}-[2 + 2] isomer has the maximum binding energy among all (C$_{20}$)$_{2}$ dimers,
this isomer should be considered a metastable configuration of 40 carbon atoms. Indeed, although $E_b$ for the
{\it open}-[2 + 2] isomer is 0.12 eV/atom higher than that of two single C$_{20}$ fullerenes, it is 0.35 eV/atom
lower than that of fullerene C$_{40}$. Therefore, it is energetically favorable for the (C$_{20}$)$_{2}$ dimer
to lose its molecular structure via the coalescence of its two C$_{20}$ clusters into one large C$_{40}$ fullerene,
by analogy with the synthesis of light nuclei [16]. In order to study the stability of the C$_{20}$ dimer against this
transformation, one should determine the energy barrier $U$ that hinders this coalescence.

Our numerical simulation of the dynamics of the {\it open}-[2 + 2] isomer at $T=2000\div 4000$ K indicates that
it does lose its molecular structure (in which its two C$_{20}$ clusters retain their identity). The average time $\tau$
of such a loss ranges from 1 ps to 10 ns depending on the temperature $T$. However, we found that the C$_{40}$
fullerene is not formed in this case but rather various C$_{40}$ clusters appear with a binding energy lower than that
of the 40-atom fullerene (but higher than that of the {\it open}-[2 + 2] isomer).

The shape of the C$_{40}$ cluster that forms most often is shown in Fig. 4a. At first glance, it forms via the rotation
of one C$_{20}$-C$_{20}$ bond through 90$^0$ (as in the Stone-Wales transformation in fullerene
C$_{60}$ [17]). However, analysis demonstrates that the character of the rearrangement of the C-C bonds is more
complex. After the 1-5 and 4-6 bonds (which are intracluster bonds for the C$_{20}$ fullerenes in the
(C$_{20}$)$_{2}$ dimer) have been broken (Fig. 2c), new bonds (1-4, 5-6) appear. Atoms 1 and 4, which belonged
earlier to different C$_{20}$ fullerenes, are {}``collectivized'', and 1-4 bond becomes an intracluster bond for the
C$_{40}$ cluster. The binding energy of this cluster is $E_b=6.25$ eV/atom, which is 0.05 eV/atom higher than that
of the {\it open}-[2 + 2] isomer. This cluster can be considered as a defect isomer of the C$_{40}$ fullerene
whose surface contains two nonagons apart from pentagons and hexagons. Recall that, in the C$_{40}$ fullerene, the C-C
bonds between the nearest carbon atoms form twelve pentagons and ten hexagons.

To find the energy barrier $U$ that hinders the transformation of the (C$_{20}$)$_{2}$ dimer into the C$_{40}$
cluster shown in Fig. 4a, we calculated the potential relief
$E_{\mathrm{pot}}\left(\left\{ \mathbf{R}_{i}\right\} \right)$ in the vicinity of the {\it open}-[2 + 2] metastable
state (Fig. 5). Thus, we found $U=2.5$ eV and determined the atomic configurations of two transient atomic states
corresponding to the saddle points of $E_{\mathrm{pot}}\left(\left\{ \mathbf{R}_{i}\right\} \right)$ and the
atomic configuration of an intermediate metastable state corresponding to a local minimum of
$E_{\mathrm{pot}}\left(\left\{ \mathbf{R}_{i}\right\} \right)$. In the intermediate state (Fig. 4b), another bond
appears between the C$_{20}$ fullerenes; the midpoint of this bond is the center of symmetry of this atomic
configuration. The intermediate state is separated from the {\it open}-[2 + 2] isomer by a relatively low barrier
($U=0.63$ eV) and is located in a relatively flat section of the potential energy surface. Therefore, numerous
configurations with energies close to the intermediate-state energy can exist; the (C$_{20}$)$_{2}$ dimer can easily
transform into these configurations by passing through barrier 2 and stay in these states for a long time until it
coalesces to form the C$_{40}$ cluster. This behavior is supported by the molecular dynamics data.

We also observed the coalescence of C$_{20}$ fullerenes into other C$_{40}$ clusters; some of them (after
relaxation) are depicted in Fig. 6. The binding energy of one of these clusters (Fig. 6a) is $E_b=6.195$ eV/atom,
which is close to that of the {\it open}-[2 + 2] isomer (the former energy is even slightly lower); that is, the transition
occurs between two almost energetically degenerate configurations. The surface of this cluster has one eight-member
and two ten-member {}``windows''. The binding energies of the other C$_{40}$ clusters are significantly higher than that
of the {\it open}-[2 + 2] isomer. The value of $E_b$ is the higher, the smaller the number of $N$-gons
with $N\geq7$ in the cluster and/or the smaller the number of atoms $N$ in them (i.e., the closer the structure of the
cluster to the C$_{40}$ fullerene). For example, $E_b=6.32$ eV/atom in a C$_{40}$ cluster with two octagons
(Fig. 6b), $E_b=6.36$ eV/atom in a C$_{40}$ cluster with one octagon and one heptagon (Fig. 6c), and
$E_b=6.49$ eV/atom in a C$_{40}$ cluster with one heptagon (Fig. 6d). We also observed transformations of the
(C$_{20}$)$_{2}$ dimer into C$_{40}$ clusters having one octagon and two heptagons ($E_b=6.34$ eV/atom),
four heptagons ($E_b=6.35$ eV/atom), and so on. All these clusters are so-called nonclassical C$_{40}$ fullerene
isomers, since, apart from pentagons and hexagons, they contain at least one $N$-gon with $N\geq 7$ [18].

An analysis of the shape of the potential energy surface shows that the heights of the energy barriers to the transitions of the
(C$_{20}$)$_{2}$ dimer into C$_{40}$ clusters of different types differ substantially; in
most cases, we have $U=2\div 4$ eV. When simulating the evolution of the (C$_{20}$)$_{2}$ dimer, we assumed
that the total energy of the system is constant. As a result, the formation of a C$_{40}$ cluster with a binding energy
$E_b$ higher than that of the (C$_{20}$)$_{2}$ dimer (i.e., with a lower potential energy $E_{pot}$) is
accompanied by heating of the cluster. This leads to the annealing of defects ($N$-gons with $N\geq 7$) and
sequential transitions of the C$_{40}$ cluster into configurations with a progressively higher binding energy.
However, the system temperature also increases. As a result, the C$_{40}$ cluster decomposes, i.e., loses its spherical
shape and transforms into quasi-one-dimensional or quasi-twodimensional configurations. Although this
decomposition is unfavorable from a thermodynamic viewpoint (since it increases the potential energy), it is,
nevertheless, irreversible. This is related to the presence of numerous low-dimensional configurations with close
energies into which the C$_{40}$ cluster transforms sequentially after decomposition. The number of such
configurations that are geometrically close to the fullerene and can transform into it is very small. Therefore, the
process of decomposition is irreversible despite the high potential energy of the decomposed cluster and the relatively
low energy barrier separating the atomic configurations appearing after decomposition from the compact fullerene.
A similar situation was detected in the simulation of the thermal stability of the C$_{20}$ fullerene [13]. Note that
we did not observe the reverse transformation from the C$_{40}$ cluster into the (C$_{20}$)$_{2}$ dimer.

The (C$_{20}$)$_{2}$ dimer loses its molecular structure not only through the coalescence of two C$_{20}$
fullerenes into a C$_{40}$ cluster. We also observed another scenario of stability loss of the {\it open}-[2 + 2]
isomer. At a high temperature, only one of the C$_{20}$ fullerenes forming the (C$_{20}$)$_{2}$ dimer can
decompose, while the other C$_{20}$ fullerene retains its shape. Figure 7 shows the typical atomic configuration
formed upon this decomposition. Its binding energy ($E_b=6.14$ eV/atom) is lower than that of the
(C$_{20}$)$_{2}$ dimer. Therefore, the decay of one C$_{20}$ fullerene is accompanied by an increase in
the potential energy $E_{pot}$ and, hence, by cooling of the cluster. The reverse transformation of the system into the
(C$_{20}$)$_{2}$ dimer does not occur for the reasons discussed above.

An analysis of the molecular dynamics data reveals that the decay of one C$_{20}$ fullerene in the
(C$_{20}$)$_{2}$ dimer can proceed in different ways, see [19] for more details. Figure 8 shows the dependence of
the potential energy $E_{pot}$ on the reaction coordinate $X$ for one of the decay channels. The sequence of the
breaking of interatomic bonds is identical to that for the decomposition of a single C$_{20}$ fullerene [13]: first, two
C-C bonds are broken simultaneously and two adjacent octagons form on the {}``lateral surface'' and, then, three more
C-C bonds are sequentially broken. As a result, the number of octagons on the lateral surface increases to five and,
finally, the defect C$_{20}$ fullerene decomposes (Fig. 7). The barrier height is $U=5.0$ eV, as in the case of a single
C$_{20}$ fullerene (Fig. 8). However, there are two differences: (i) $E_{pot}(X)$ reaches a maximum as the fifth
rather than the fourth C-C bond is broken, and (ii) the atomic configuration with two broken C-C bonds is not
metastable. Although there exist other channels of decomposition of one C$_20$ fullerene in the (C$_{20}$)$_{2}$
dimer (they are characterized by $U=3\div 5$ eV), the decomposition usually begins with the breakage of two C-C
bonds, as in the case described above.

The energy barrier to the coalescence of two C$_{20}$ fullerenes
into a C$_{40}$ cluster ($U=2\div 4$ eV) is somewhat lower than
the barrier to the decomposition of one C$_{20}$ fullerene
($U=3\div 5$ eV). Therefore, as the temperature decreases, the
former mechanism of stability loss in the (C$_{20}$)$_{2}$ dimer
becomes more important. It should be noted, however, that there
are cases that a clear distinction  between these two decay
channels cannot be made. For example, we observed a situation in
which a C$_{40}$ cluster formed before the decomposition of one
C$_{20}$ fullerene was completed. In another case, in contrast, a
C$_{40}$ cluster existed only for a short time (less than 1 ps)
and then decomposed in such a manner that only one of its halves
transformed into a low-dimensional configuration, whereas the
other half retained the shape of the C$_{20}$ fullerene. The
latter scenario occured primarily when the C$_{40}$ cluster formed
at the first stage was a defect C$_{40}$ fullerene isomer similar
to that shown in Fig. 6a. With these exceptions, the character of
the molecular structure loss of the (C$_{20}$)$_{2}$ dimer (i.e.,
coalescence or decomposition) is determined unambiguously.

\vskip 5mm

\centerline{\bf 5. CONCLUSIONS}

In this study, we have shown that there is no energy barrier to the formation of a (C$_{20}$)$_{2}$ cluster dimer
from two single C$_{20}$ fullerenes; that is, the formation of this dimer ({\it open}-[2 + 2] isomer) is favorable
from both the kinetic and thermodynamic viewpoint. This result was obtained by the tight-binding method and
supplements the data obtained earlier [8] using the Hartree-Fock and density functional methods.

The barriers that hinder the molecular structure loss of the
(C$_{20}$)$_{2}$ dimer are $U=2\div 4$ eV for the coalescence of
two C$_{20}$ fullerenes into a C$_{40}$ cluster and $U=3\div 5$ eV
for the decomposition of one C$_{20}$ fullerene in the
(C$_{20}$)$_{2}$ dimer. Thus, although the stability of the
(C$_{20}$)$_{2}$ dimer is lower than that of the C$_{20}$
fullerene (for which $U=5$ eV), it is rather high. In the near
future, it would be interesting to study and simulate the
potential energy surfaces and dynamics of the
C$_{20}$-fullerene-based three-dimensional structures discussed in
the literature, determine their stability, and calculate the
formation energies of various structural defects.

\newpage

\includegraphics[width=\hsize,height=15cm]{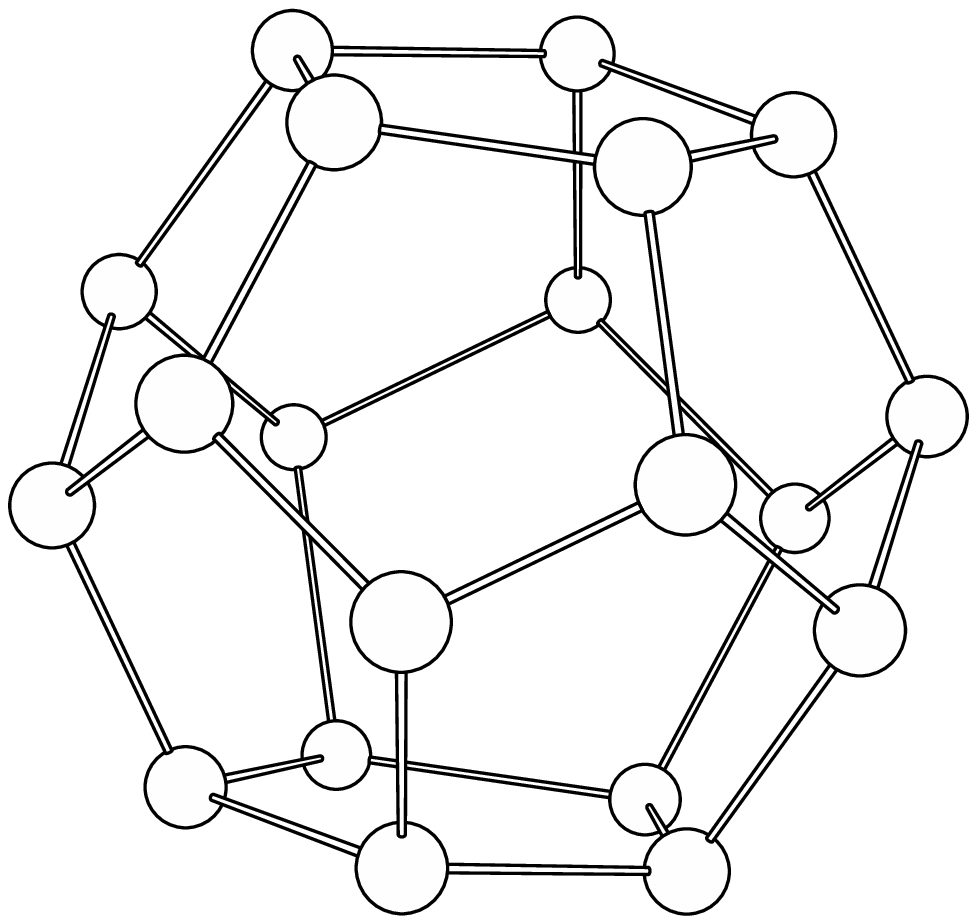}

\vskip 5mm

Fig. 1. Fullerene C$_{20}$. The binding energy is $E_b=6.08$ eV/atom.

\newpage

\includegraphics[width=11cm,height=5cm]{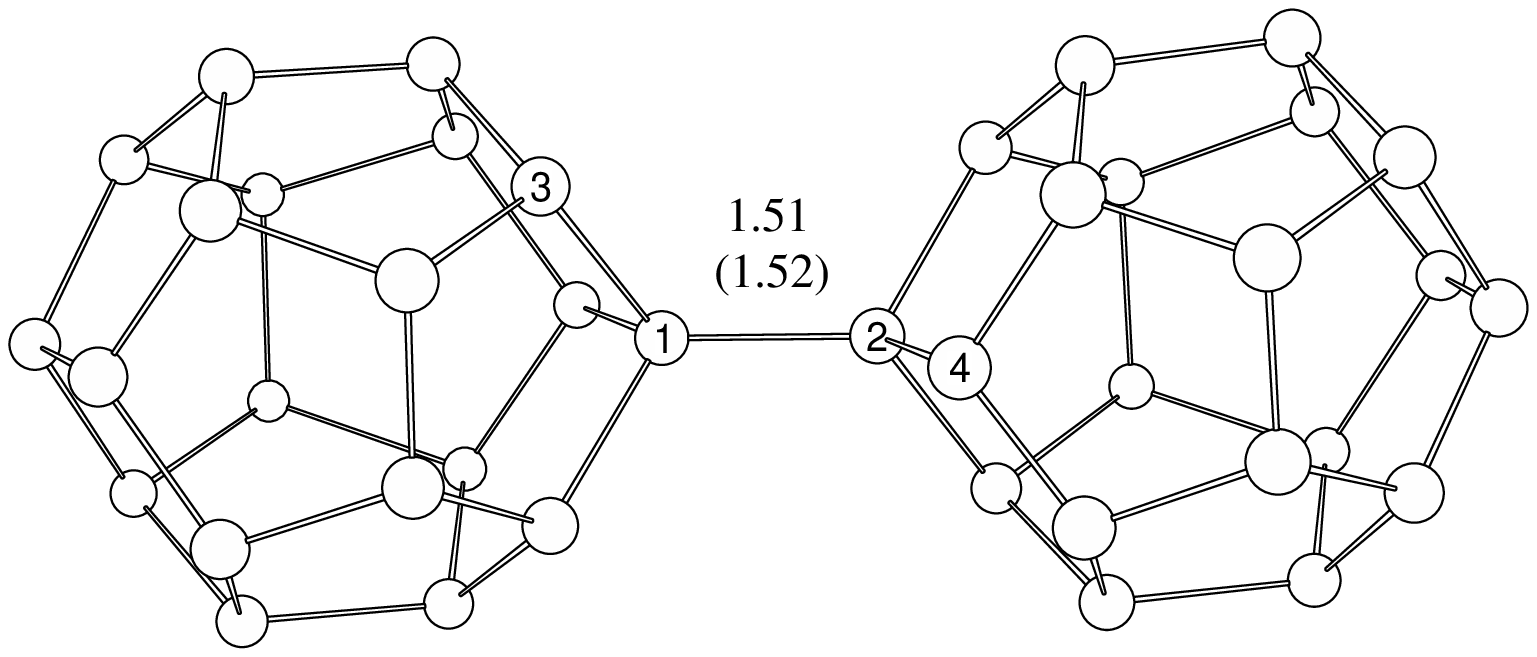}
\vskip 5mm
\includegraphics[width=11cm,height=5cm]{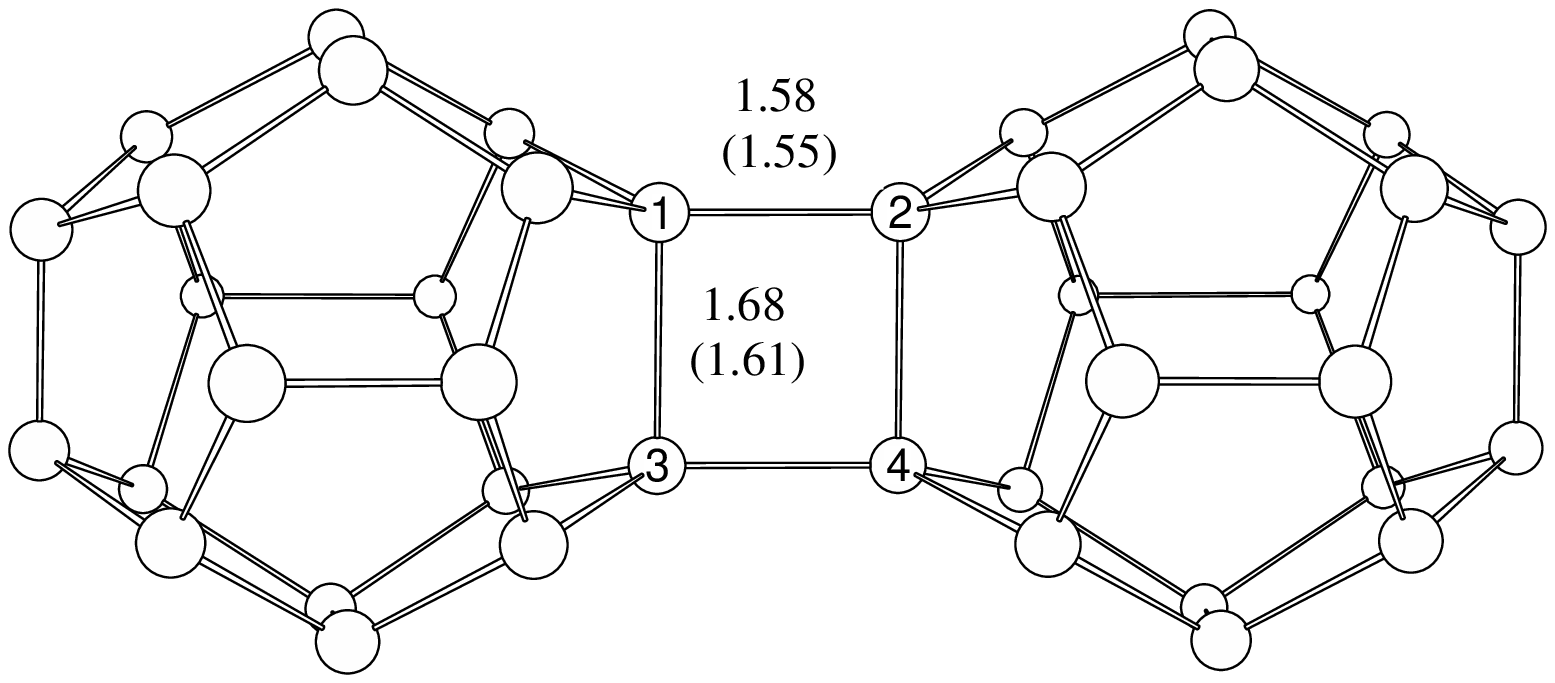}
\vskip 5mm
\includegraphics[width=11cm,height=5cm]{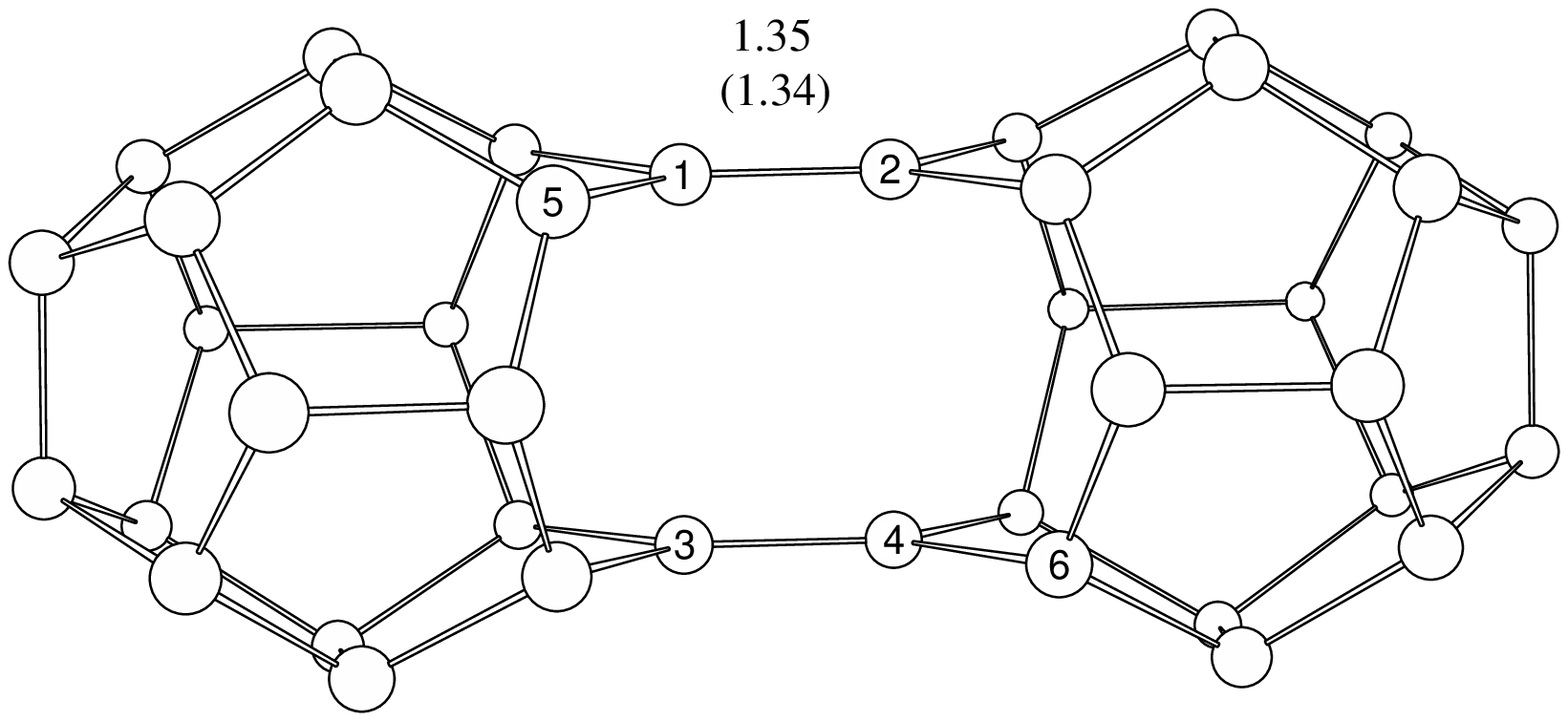}

\vskip 5mm

Fig. 2. Isomers (C$_{20}$)$_{2}$: (a) [1+1] with a binding energy $E_b=$ 6.14 eV/atom,
(b) [2+2] with a binding energy $E_b=$ 6.16 eV/atom, and (c) {\it open}-[2+2] with a binding energy
$E_b=$ 6.20 eV/atom. The numerals above parentheses indicate the bond lengths in angstroms, and the numerals in
parentheses are the bond lengths calculated in [8] using the density functional method.

\newpage

\includegraphics[width=7cm,height=5cm]{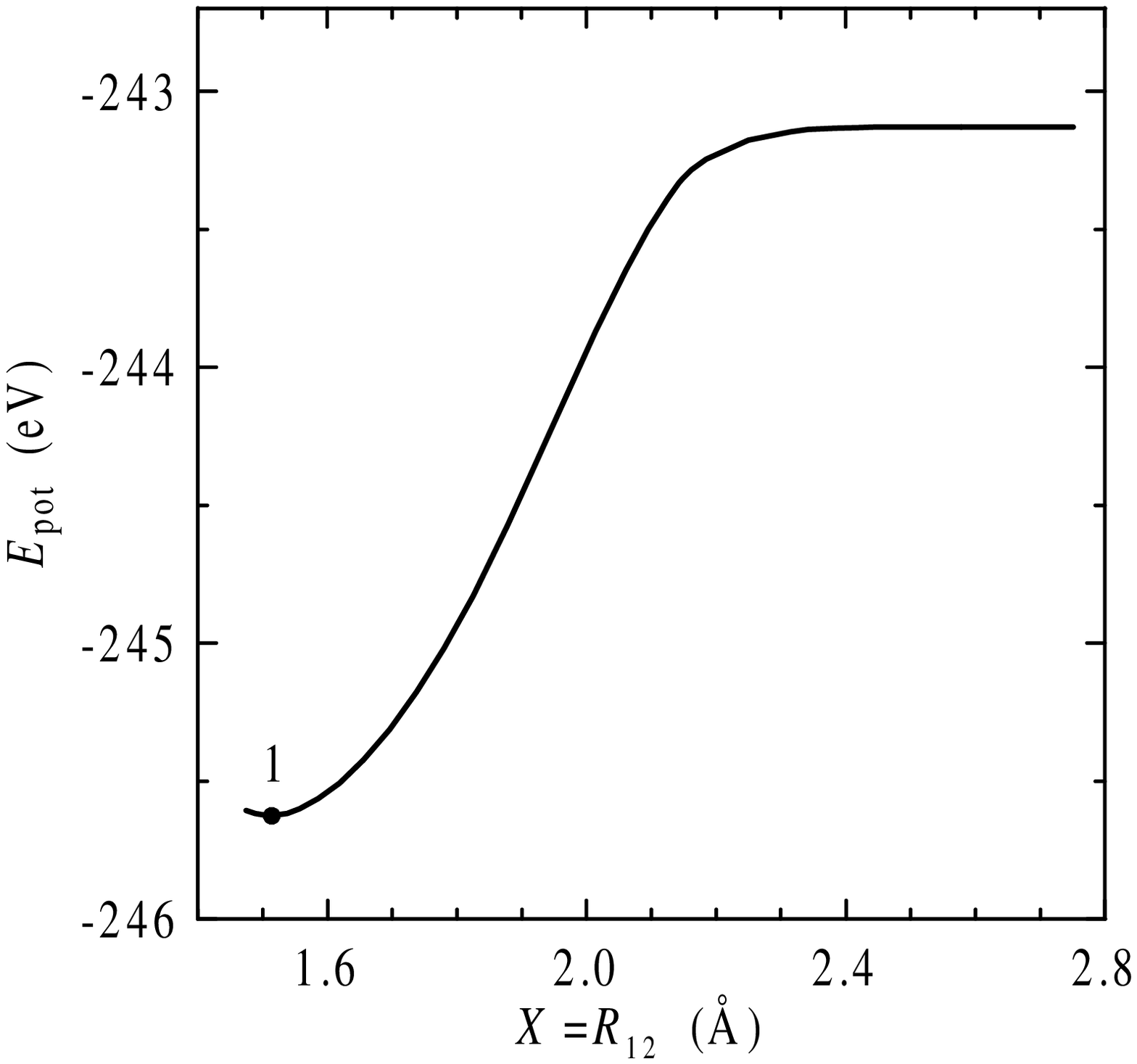}
\vskip 5mm
\includegraphics[width=7cm,height=5cm]{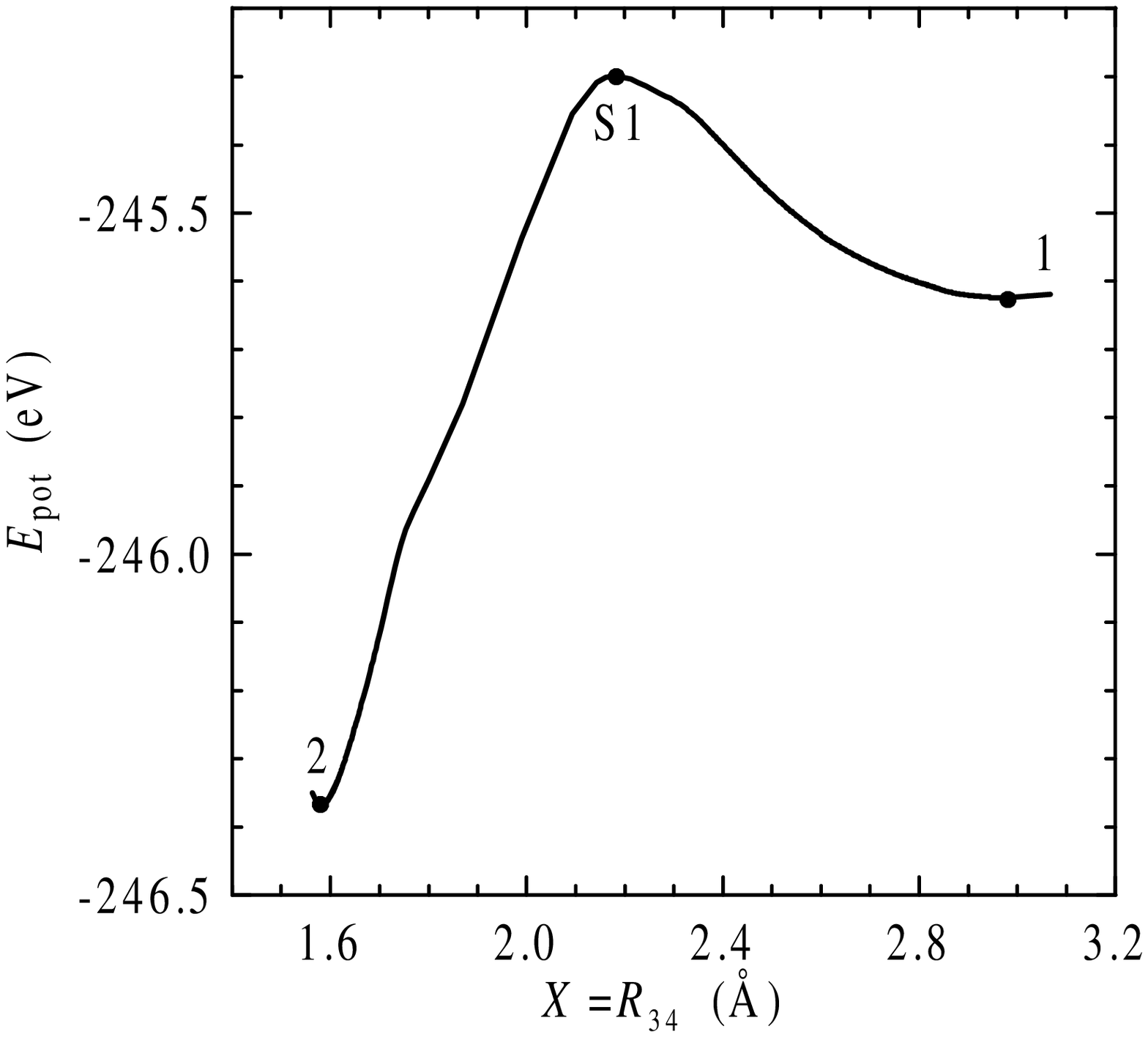}
\vskip 5mm
\includegraphics[width=7cm,height=5cm]{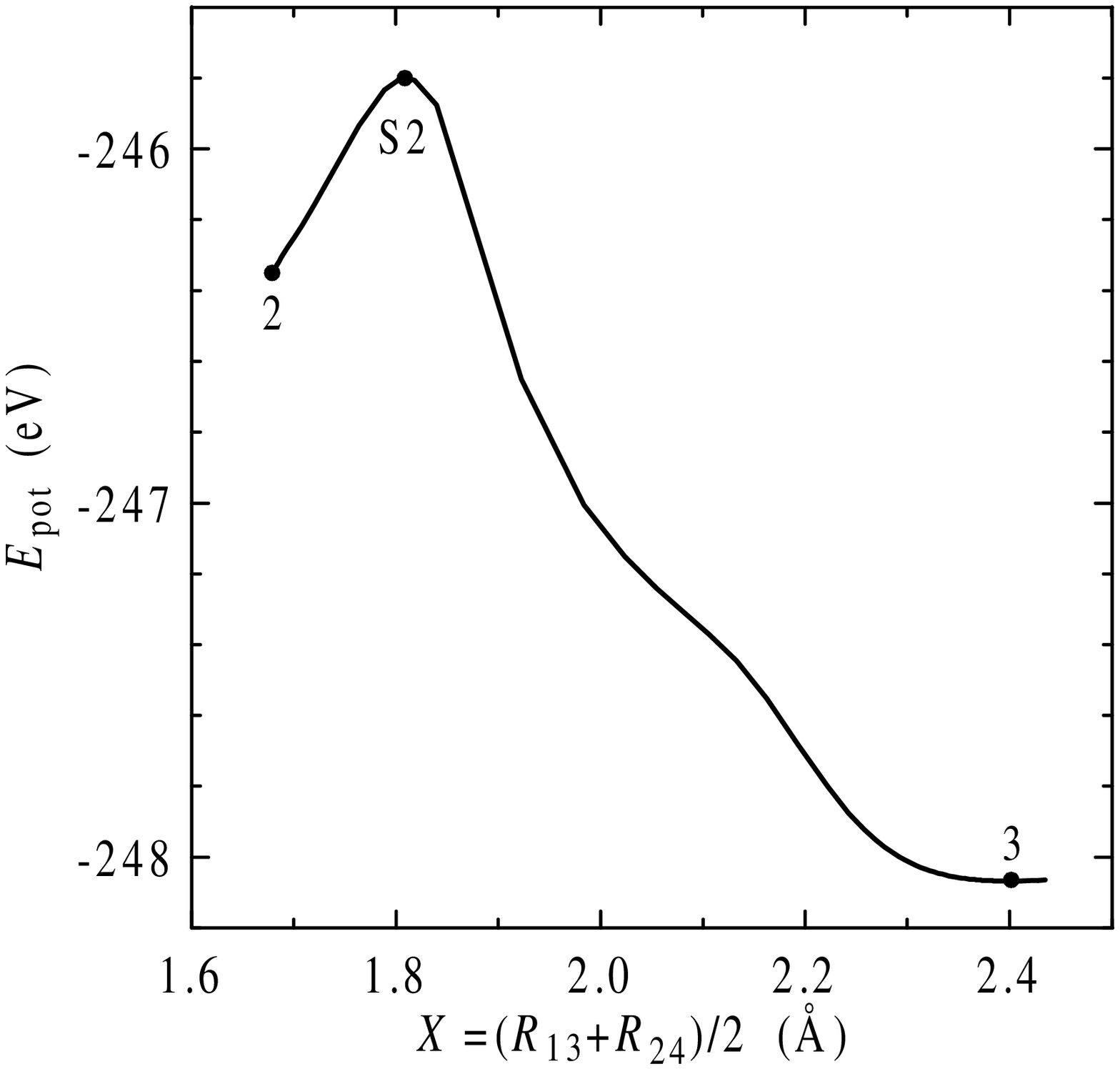}

Fig. 3. Dependence of the potential energy $E_{pot}$ of the
(C$_{20}$)$_{2}$ dimer on the reaction coordinate $X$ for (a) the
C$_{20}$+C$_{20}$ $\rightarrow$ [1 + 1], (b) [1 + 1] $\rightarrow$
[2 + 2], and (c) [2 + 2] $\rightarrow$ {\it open}-[2 + 2]
transitions. Points in curves correspond to (1) the [1 + 1]
isomer, (2) [2 + 2] isomer, and (3) {\it open}-[2 + 2] isomer. S1
and S2 are the $E_{pot}(X)$ maxima, i.e., saddle points in
$E_{\mathrm{pot}}\left(\left\{\mathbf{R}_{i}\right\} \right)$. The
reference point was taken to be the energy of 40 isolated carbon
atoms. The $X$ coordinates are chosen in terms of interatomic
distances (see Fig. 2).

\newpage

\includegraphics[width=11cm,height=5cm]{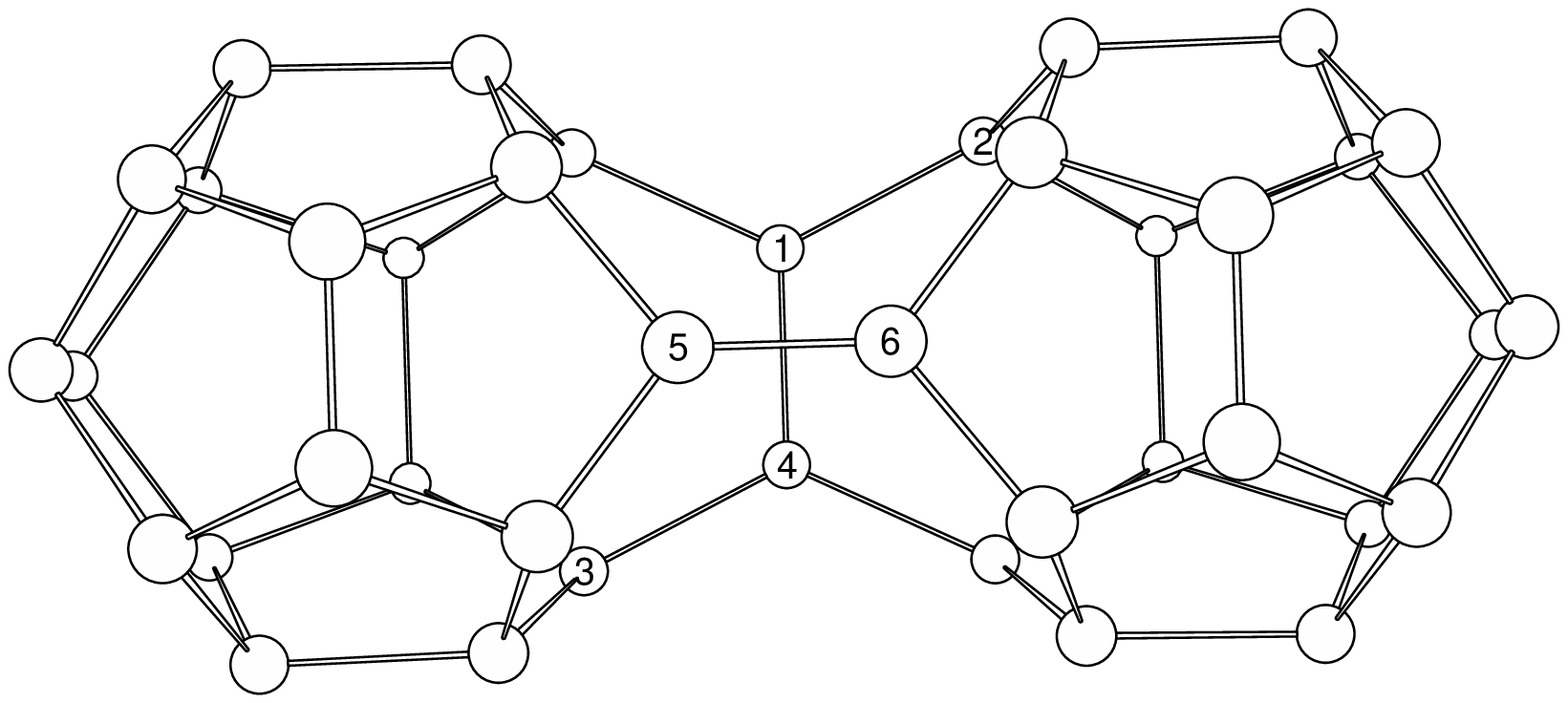}
\vskip 5mm
\includegraphics[width=11cm,height=5cm]{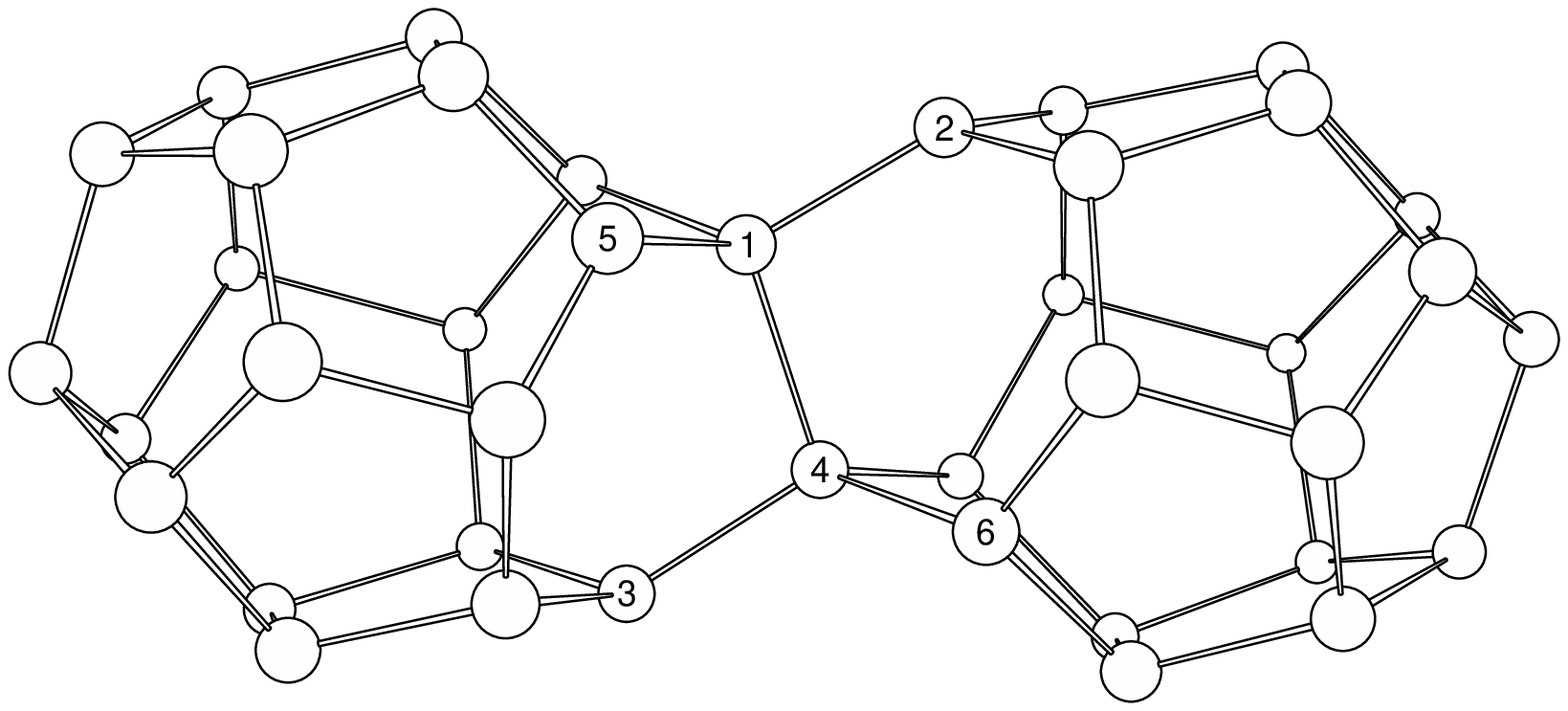}

\vskip 5mm

Fig. 4. (a) One of the C$_{40}$ clusters formed upon the
coalescence of the C$_{20}$ fullerenes in the (C$_{20}$)$_{2}$
dimer. The binding energy is $E_b=6.25$ eV/atom. (b) The
intermediate metastable state that forms when the {\it open}-[2 +
2] isomer transforms into a C$_{40}$ cluster (see Fig. 5).
$E_b=6.19$ eV/atom. The numbering of atoms is identical to that in
Fig. 2c.

\newpage

\includegraphics[width=13cm,height=12cm]{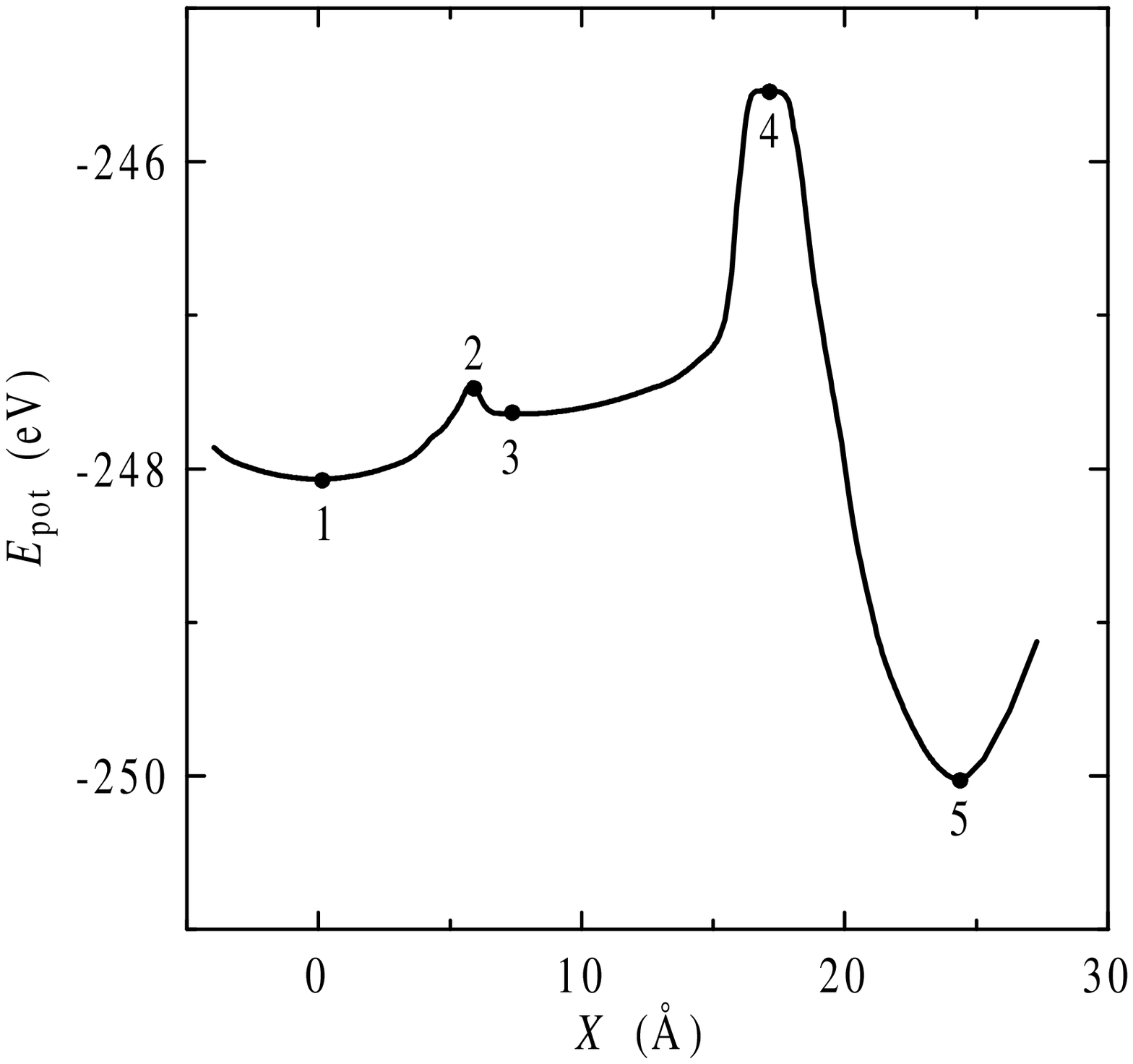}

\vskip 5mm

Fig. 5. Dependence of the potential energy $E_{pot}$ of the
(C$_{20}$)$_{2}$ dimer on the reaction coordinate $X$ for the
transformation of the {\it open}-[2 + 2] isomer into the C$_{40}$
cluster shown in Fig. 4a. Points in the curve indicate (1) the
{\it open}-[2 + 2] isomer; (2, 4) local maxima in $E_{pot}(X)$,
i.e. saddle points for
$E_{\mathrm{pot}}\left(\left\{\mathbf{R}_{i}\right\} \right)$; (3)
the local minimum in $E_{pot}(X)$ corresponding to the metastable
intermediate state shown in Fig. 4b; and (5) a C$_{40}$ cluster
(Fig. 4a). The reference point is taken to be the energy of 40
isolated carbon atoms. The reaction coordinate is chosen to be the
length of the path that passes through the corresponding saddle
point in the $(3n-6)$-dimensional space and connects the {\it
open}-[2 + 2] isomer and the C$_{40}$ cluster (as in [14]).

\newpage

\includegraphics[width=9cm,height=4.5cm]{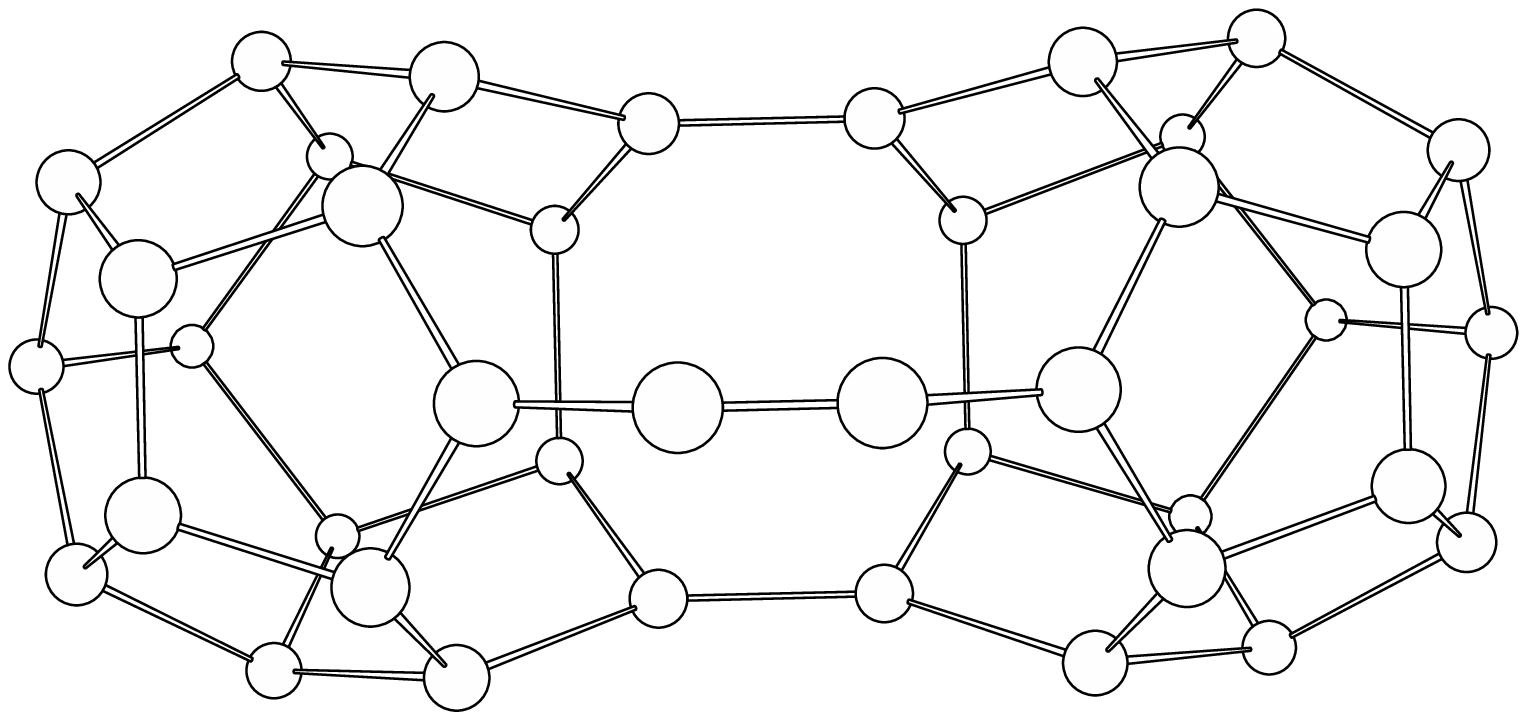}
\vskip 5mm
\includegraphics[width=9cm,height=4.5cm]{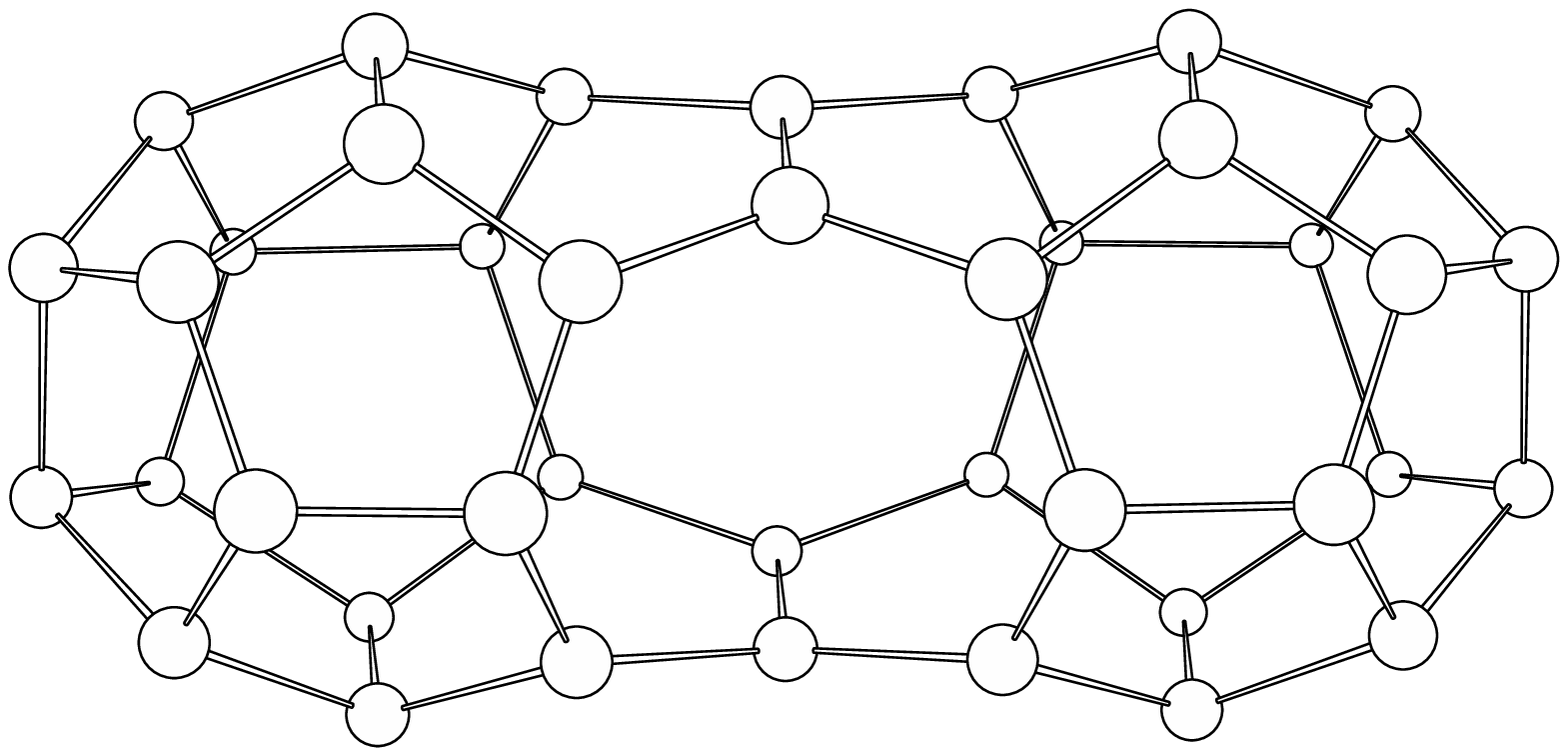}
\vskip 5mm
\includegraphics[width=9cm,height=4.5cm]{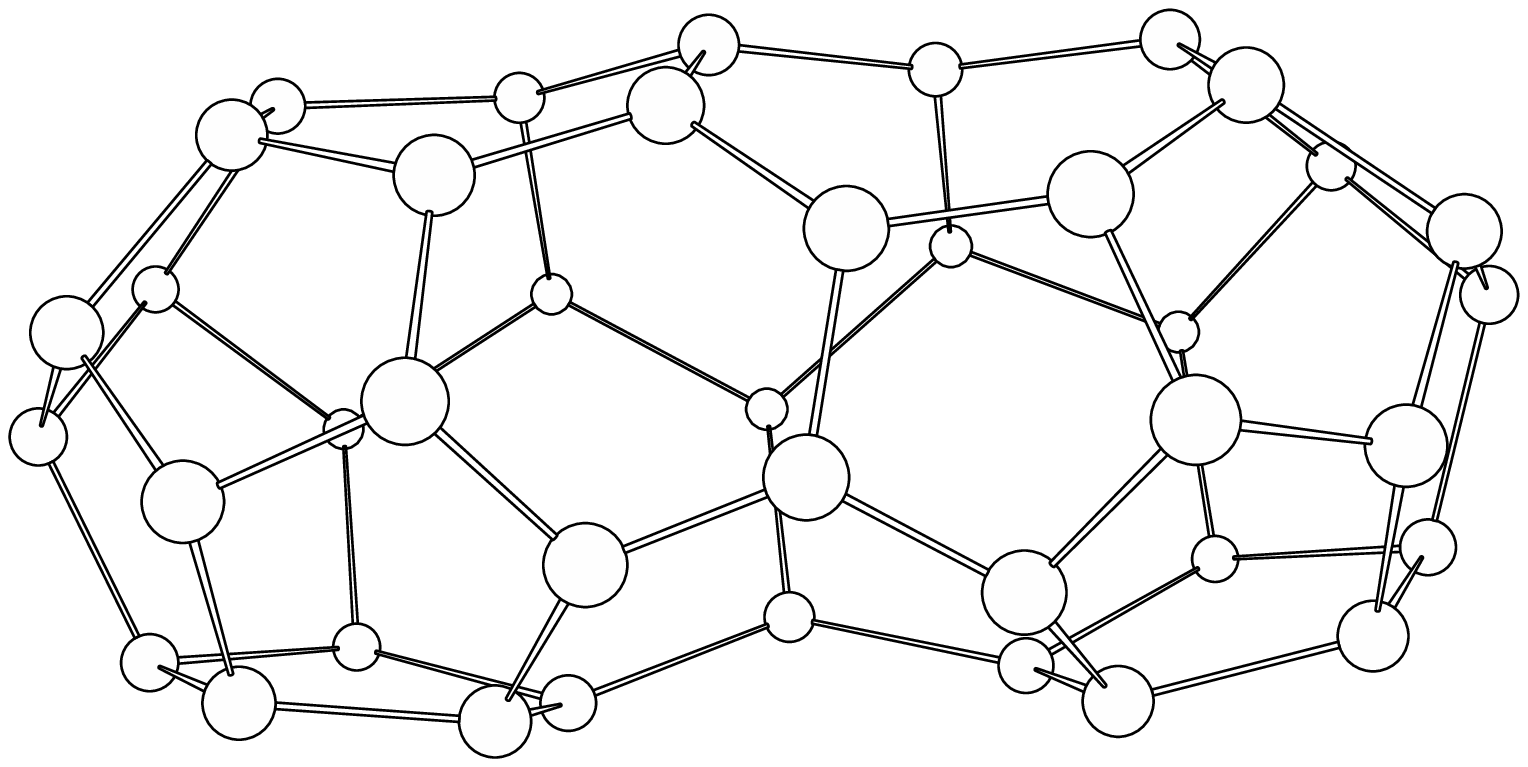}
\vskip 5mm
\includegraphics[width=8cm,height=4.5cm]{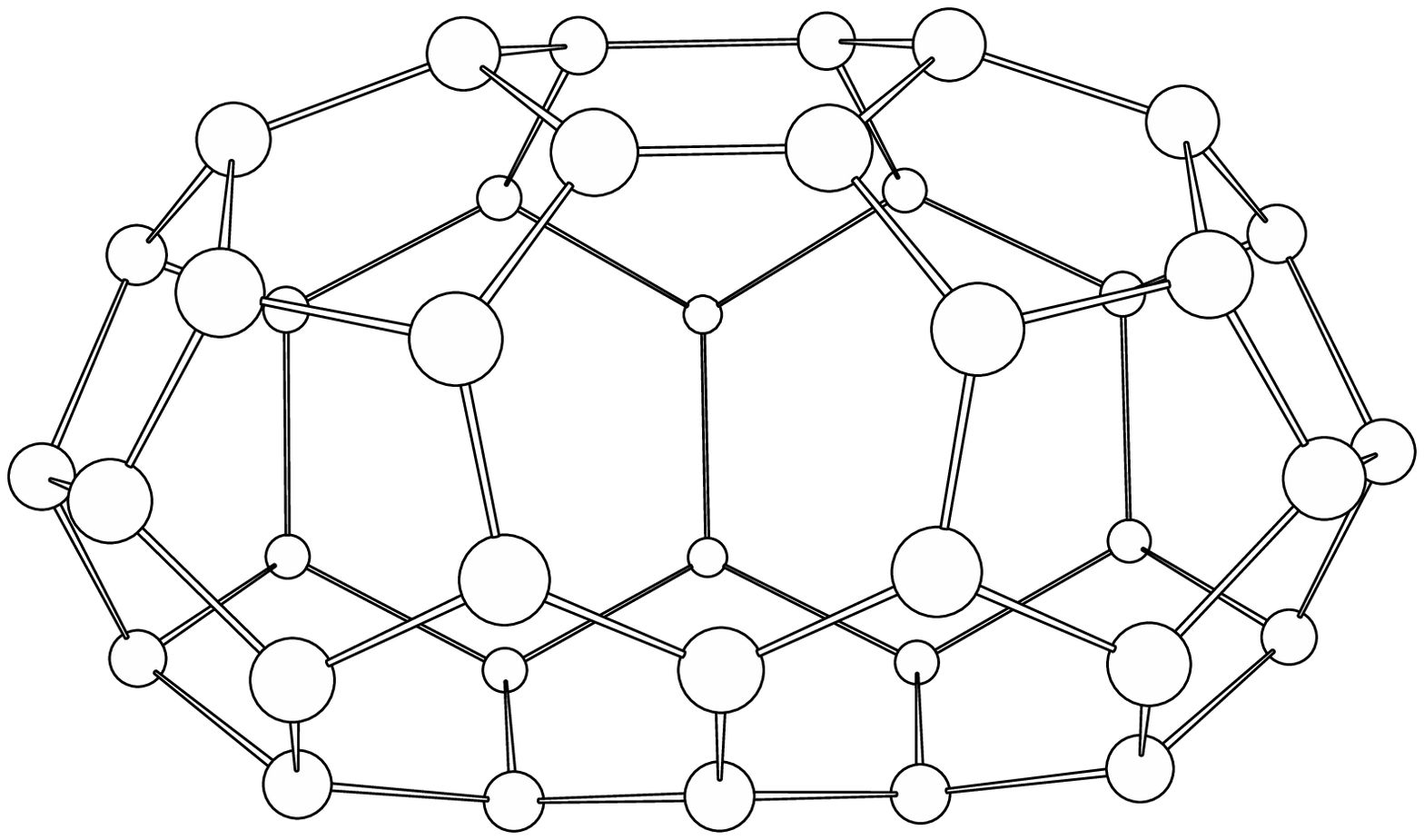}

\vskip 5mm

Fig. 6. Some of the C$_{40}$ clusters formed from the (C$_{20}$)$_{2}$ dimer at a high temperature. Their
binding energies $E_b$ (after relaxation) are (a) 6.195, (b) 6.32, (c) 6.36, and (d) 6.49 eV/atom.

\newpage

\includegraphics[width=13cm,height=10cm]{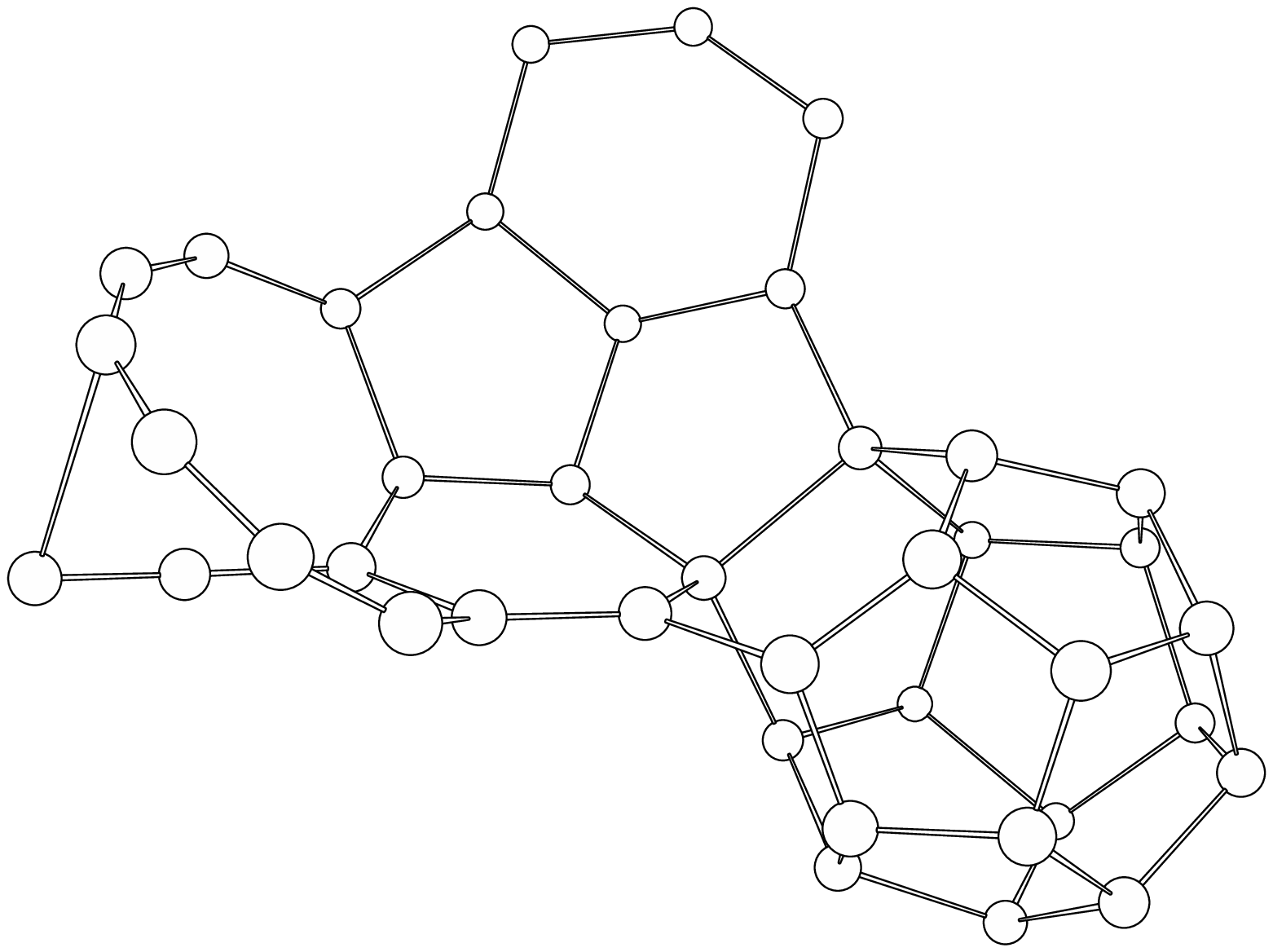}

\vskip 5mm

Fig. 7. Atomic configuration after the decomposition of one C$_{20}$ fullerene in the (C$_{20}$)$_{2}$ dimer.
The binding energy is $E_b=6.14$ eV/atom.

\newpage

\includegraphics[width=13cm,height=12cm]{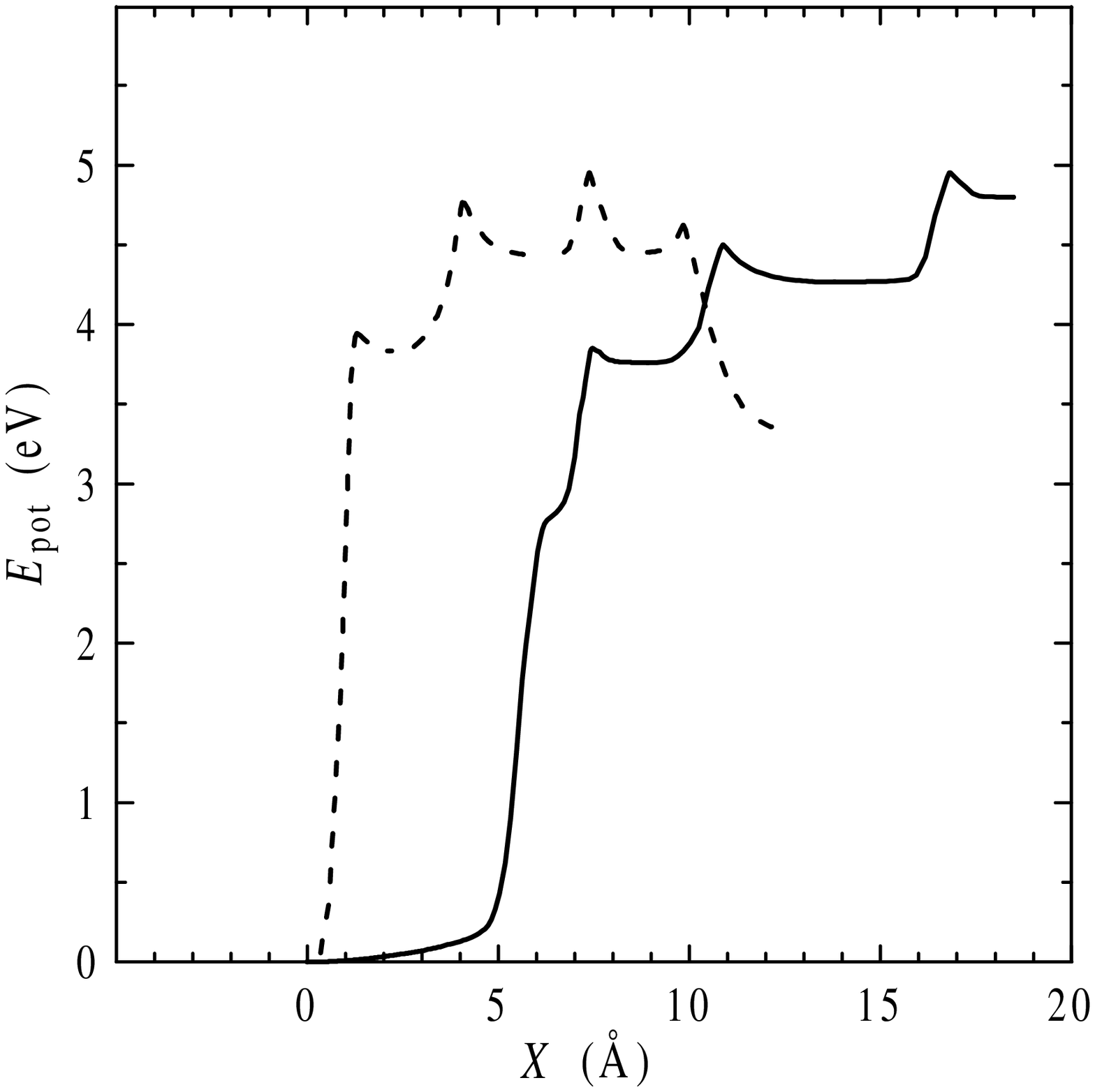}

\vskip 5mm

Fig. 8. Dependence of the potential energy $E_{pot}$ of the (C$_{20}$)$_{2}$ dimer on the reaction coordinate
$X$ for the decay of one C$_{20}$ fullerene in the (C$_{20}$)$_{2}$ dimer (solid line) and for the decay of a
single C$_{20}$ fullerene (dashed line). The reference point is taken to be the energy of the {\it open}-[2 + 2] isomer
and the C$_{20}$ fullerene, respectively. The reaction coordinate $X$ is the same as in Fig. 5.

\end{document}